\documentclass[longauth]{aa} 
\usepackage{graphicx}
\usepackage{txfonts}
\begin{document}
   \title{MAGIC upper limits to the VHE gamma-ray flux of 3C454.3 in
   high emission state}

\author{
 H.~Anderhub\inst{1} \and
 L.~A.~Antonelli\inst{2} \and
 P.~Antoranz\inst{3} \and
 M.~Backes\inst{4} \and
 C.~Baixeras\inst{5} \and
 S.~Balestra\inst{3} \and
 J.~A.~Barrio\inst{3} \and
 H.~Bartko\inst{6} \and
 D.~Bastieri\inst{7} \and
 J.~Becerra Gonz\'alez\inst{8}
% ...
}

\offprints{F. Tavecchio, N. Mankuzhiyil and V. Scapin}

\institute {ETH Zurich, CH-8093 Switzerland
 \and INAF National Institute for Astrophysics, I-00136 Rome, Italy
 \and Universidad Complutense, E-28040 Madrid, Spain
 \and Technische Universit\"at Dortmund, D-44221 Dortmund, Germany
 \and Universitat Aut\`onoma de Barcelona, E-08193 Bellaterra, Spain
 \and Max-Planck-Institut f\"ur Physik, D-80805 M\"unchen, Germany
 \and Universit\`a di Padova and INFN, I-35131 Padova, Italy
 \and Inst. de Astrof\'{\i}sica de Canarias, E-38200 La Laguna, Tenerife, Spain
% ...
}

\date{Received ... ; accepted ...}

\abstract 
{}  
{We report upper limits to the very high energy flux ($E>100$ GeV) of
the flat spectrum radio quasar 3C454.3 ($z=0.859$) derived by the
Cherenkov telescope MAGIC during the high states of July/August and
November/December 2007. We compare the upper limits derived in both
time slots with the available quasi-simultaneous MeV-GeV data from the
AGILE $\gamma$-ray satellite and interpret the observational results
in the context of leptonic emission models.}
{The source was observed with the MAGIC telescope during the active
phases of July-August 2007 and November-December 2007 and the data
were analyzed with the MAGIC standard analysis tools. For the periods
around the ends of July and November, characterized by the most
complete multifrequency coverage, we constructed the spectral energy
distributions using our data together with nearly simultaneous
multifrequency (optical, UV, X-ray and GeV) data.}
{Only upper limits can be derived from the MAGIC data. The upper
limits, once corrected for the expected absorption by the
extragalactic background light, together with nearly simultaneous
multifrequency data, allow us to constrain the spectral energy
distribution of 3C454.3. The data are consistent with the model
expectations based on the inverse Compton scattering of the ambient
photons from the broad line region by relativistic electrons, which
robustly predicts a sharp cut-off above 20-30 GeV.}
{}

   \keywords{quasars: individual (3C454.3) -- Gamma rays: observations --
Gamma rays: theory}

\titlerunning{MAGIC upper limits to 3C454.3 in high state}

\maketitle

\section{Introduction}

The present list of known extragalactic sources of Very High Energy
(VHE; defined here as $E>100$ GeV) radiation includes 24
sources\footnote{25, pending the confirmation of the radiogalaxy 3C66B
detected by MAGIC (Aliu et al. 2008).} (e.g. Aharonian et al. 2008a,
De Angelis et al. 2008)\footnote{see also {\tt
http://www.mppmu.mpg.de/$\sim$rwagner/sources}}. As expected, the
majority of these sources (18) belong to the high-peaked BL Lac
class. The remaining 6 are four low peaked-BL Lac objects (BL Lac:
Albert et al. 2007; W Comae: Acciari et al. 2008; S5 0716+71: Teshima
et al. 2008; 3C66A: Swordy et al. 2008), a radiogalaxy (M87: Aharonian
et al. 2003, 2006a) and a quasar (3C279: Albert et al. 2008a).

Although the detection of 3C279 indicates that also quasars can, to
some extent, emit VHE radiation, general theoretical arguments support
the view that, due to internal absorption (e.g. Liu \& Bai 2006,
Reimer 2007) and/or to the decrease of the cross section for inverse
Compton scattering (e.g. Tavecchio \& Ghisellini 2008) powerful Flat
Spectrum Radio Quasars (FSRQs) cannot be prominent VHE
emitters. Moreover, FSRQs are generally located at relatively high
redshifts, implying a huge absorption of $\gamma$-rays by the
Extragalactic Background Light (EBL). On the other hand, the detection
of these sources at VHE would be important for our understanding of
their structure and of acceleration/emission mechanisms and would
provide a unique opportunity to probe the EBL at relatively high
redshifts, allowing to study its evolution over cosmic time.

3C454.3 ($z=0.859$) is a well known FSRQ, detected several times in
the $\gamma$-ray band by the EGRET telescope onboard {\it CGRO}, with
an average photon index of $\Gamma=2.2$ (Hartmann et al. 1999). In
2005 it underwent a very active phase in optical and X-ray bands,
triggering intensive observations in the radio, optical and X-ray
({\it Swift}, {\it Chandra}, {\it INTEGRAL}) bands (Villata at
al. 2006, Giommi et al. 2006, Pian et al. 2006). Unfortunately no
$\gamma$-ray satellite was operating in the GeV domain at that time
and no information was obtained in that band.

During the summer of 2007, 3C454.3 was active again, reaching a level
of the optical emission comparable to that of 2005. Several observations in
the optical, X-ray and $\gamma$-ray band were triggered (optical: KVA,
optical-UV: {\it Swift}/UVOT, X-ray: {\it Swift}/XRT, GeV band:
AGILE/GRID). The AGILE satellite (Tavani et al. 2008), still in its
science verification phase, detected intense emission from 3C454.3
(Vercellone et al. 2008a).

Triggered by these observations, the Major Atmospheric Gamma-ray
Imaging Cherenkov (MAGIC) Telescope observed 3C454.3 in July and
August 2007. Another $\gamma $-ray active phase was recorded by AGILE
in November-December 2007 (Vercellone et al. 2008b, Vercellone et
al. 2009, in prep), which triggered further observations with MAGIC
during that period.  In the following (Section \ref{analysis}) we
describe the MAGIC observations and the analysis procedure. In
Sect. \ref{discussion} we interpret the results in the framework of
the widely assumed Synchrotron Self + External Compton (e.g. Sikora et
al. 1994) model.

\section{MAGIC observations and data analysis}
\label{analysis}

MAGIC (Baixeras et al. 2004, Cortina et al. 2005) is a new generation
Imaging Atmospheric Cherenkov Telescope at La Palma, Canary Islands,
Spain (28.3$^{\circ}$N, 17.8$^{\circ}$W, 2240~m asl). Thanks to its
low energy trigger threshold of 60~GeV, MAGIC is well suited for
multiwavelength observations together with the instruments operating
in the GeV range.  The parabolically-shaped reflector, with its total
mirror area of 236~m$^2$ allows MAGIC to sample a part of the
Cherenkov light pool and focus it onto a multi-pixel camera, composed
of 576~ultra-sensitive photomultipliers.  The total field of view of
the camera is 3.5$^\circ$ and the collection area is of the order of
$10^5$~m$^2$ at 200 GeV for a source close to zenith.

The incident light pulses are converted into optical signals,
transmitted, via optical fibers and digitized by 2-GHz flash ADCs
(Goebel et al. 2007).  The primary particle energy and incoming
direction are reconstructed done by studying the intensity (and area)
of the elliptical images and their orientation in the camera.  In
particular, the shape of the image allows to suppress hadron-induced
showers and thus to reject the hadronic background.

In July and August 2007 observations were carried out in the ON/OFF
mode, in which the source was observed on axis (for a total of 9.6
hours), while for background estimation, observations (for a total of
7.3 hours) from a region of similar conditions in the sky from
where no gamma rays are expected were used.
Later, in November and December 2007, additional observations were
performed in the false-source tracking (wobble) mode (Fomin et
al. 1994) in which the telescope was pointed alternatingly for 20
minutes to two opposite sky positions at 0.4$^{\circ}$ offset from the
source (total 6.8 hours). This procedure allows to simultaneously
determine the background and thus no extra OFF observations are
needed. The zenith angle of all these observations
ranged from 12 to 30 degrees. The weather conditions in July were not
as good as those of August, hence the event rate was lower.

The analysis was performed using the standard MAGIC analysis software
(Bretz et al. 2005).  After calibration and image cleaning based
on a two-level tail cut (6 photoelectrons for image core
and 3 photoelectrons for boundary pixels; see Fegan 1997), the camera
images were parameterized by the so-called Hillas image parameters
(Hillas 1985).  Two additional parameters, namely the time gradient
along the main shower axis and the time spread of the shower pixels,
were computed (Albert et al. 2008d). Hadronic background suppression
was achieved using the Random Forest (RF) method (Breiman 2001, Albert
et al. 2008c), in which for each event the so-called hadronness is
computed, based on the Hillas and the time parameters. The
``hadronness'' parameter can be calculated for every event, and is a
measure of the probability that the event is not $\gamma$ like.  The
RF method was also used for the energy estimation. Crab Nebula data
from the same periods and zenith angle distributions were studied
using the same analysis chain to check the validity of the results.

Since there was no significant signal found, upper limits (95\% CL)
were calculated (Rolke et al. 2005) taking into account a 30\%
systematic error in energy determination and effective area
calculation (see Albert et al. 2008b). Table~\ref{tab:ulJulyAugust}
shows the results for the July-August observations, whereas upper limits
for the November-December observation are given in 
Table~\ref{tab:ulwobble}.

\begin{table}[htdp]
\begin{center}
\begin{tabular}{|c|c|c|c|c|c|}
\hline
$\langle E\rangle$ & \multicolumn{5}{|c|}{\textbf{U.L. July 17-20 2007}}\\
\textbf{[GeV]} & N$_{\rm ON}$ &N$_{\rm OFF}$& Sign. ($\sigma$) & C.U.  & [erg cm$^{-2}$ s $^{-1}$]\\
\hline
 83 & 54188 & 54705 & -1.56 &0.04 & $0.78\times 10^{-11}$\\
\hline
 186 & 976 & 965 & 0.25 & 0.05 & $0.62\times 10^{-11}$\\
\hline
 476 & 62 & 52.3 & 0.91 & 0.03 & $0.169\times 10^{-11}$\\
\hline
\hline
$\langle E\rangle$ & \multicolumn{5}{|c|}{\textbf{U.L. August 9-22 2007}}\\
\textbf{[GeV]} & N$_{\rm ON}$ &N$_{\rm OFF}$& Sign. ($\sigma$) & C.U.  & [erg cm$^{-2}$ s $^{-1}$]\\
\hline
128 & 5453 & 5539.7 & -0.82 & 0.14 & $2.0\times 10^{-11}$\\
\hline
186 & 3892 & 3885.1 & 0.078 & 0.03 & $0.3\times 10^{-11}$\\
\hline
476 & 202 & 220.8 & -0.91 & 0.01 & $0.09\times 10^{-11}$ \\
\hline
\end{tabular}
\end{center}
\caption{Derived upper limits on flux for the July and the August 2007
data. The columns represent respectively: the average true energy, the
number of ON source events , number of background (OFF) events, the
significance, the flux upper limit in Crab Units (C.U.) and in
absolute flux units of [erg~cm$^{-2}$ s$^{-1}$].}
\label{tab:ulJulyAugust}
\end{table}%

\begin{table}[htdp]
\begin{center}
\begin{tabular}{|c|c|c|c|c|c|}
\hline
$\langle E\rangle$ & \multicolumn{5}{|c|}{\textbf{U.L. Nov.27,30 \& Dec. 1 2007}}\\
\textbf{[GeV]} & N$_{\rm ON}$ &N$_{\rm OFF}$& Sign. ($\sigma$) & C.U.  & [erg cm$^{-2}$ s $^{-1}$]\\
\hline
\hline
 113 & 39900 & 39920 & -0.07 & 0.3 & $4.6\times 10^{-11}$\\
\hline
 235 & 385& 367& 0.66 & 0.09 & $0.9 \times 10^{-11}$\\
\hline
\end{tabular}
\end{center}
\caption{Derived upper limits on flux for the November and the
December 2007 data. The columns represent respectively: the average
true energy, the number of ON source events, number of background
(OFF) events, the significance, the flux upper limit in Crab Units
(C.U.) and in absolute flux units of [erg~cm$^{-2}$ s$^{-1}$].}
\label{tab:ulwobble}
\end{table}%

\section{Discussion}
\label{discussion}    

\begin{figure*}  
   \centering
   \includegraphics[width=14 truecm, height=10. truecm]{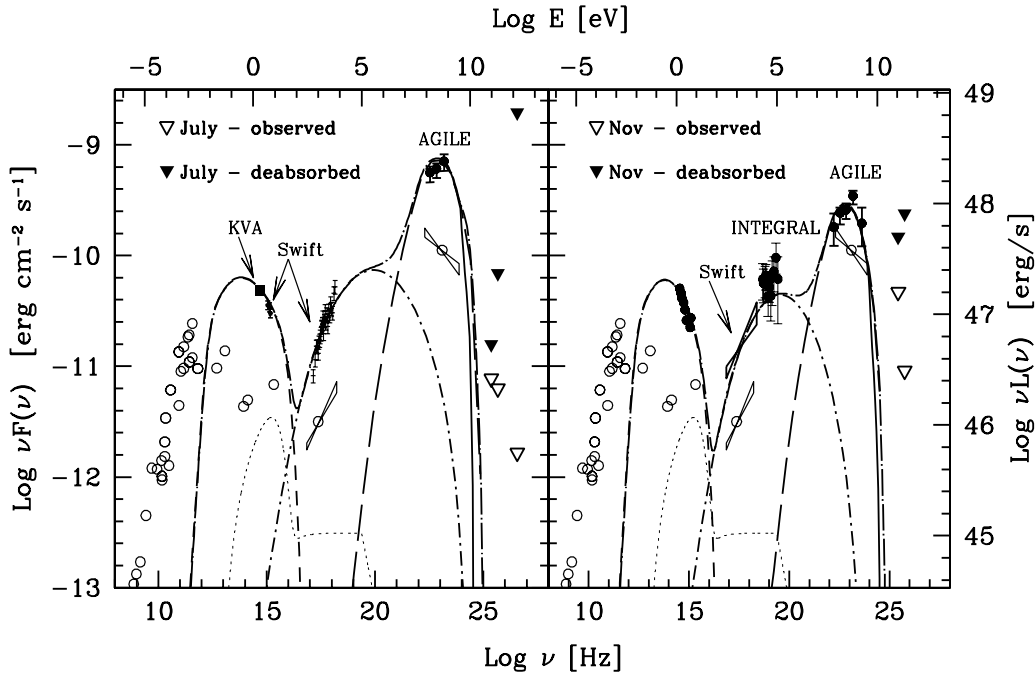}
      \caption{SED of 3C454.3 assembled with multifrequency
              information available for the period close to the MAGIC
              observation at the end of July 2007 (left panel;
              optical: KVA, optical-UV: {\it Swift}/UVOT, X-ray: {\it
              Swift}/XRT, GeV band: AGILE/GRID) and November 2007
              (right panel: optical-UV: {\it Swift}/UVOT, X-ray: {\it
              Swift}/XRT and INTEGRAL, GeV band:
              AGILE/GRID). Triangles report the observed (empty) and
              the deabsorbed (filled) upper limits of MAGIC in three
              different bands. For comparison we also report (open
              circles) historical data (Kuhr et al. 1981, NED, Gear et
              al. 1994, Stevens et al 1994, Impey \& Neugebauer 1988,
              Smith et al. 1988 for radio and optical; Tavecchio et
              al. 2007 for X-rays from {\it Chandra}). The open circle and
              the bow-tie in the MeV-GeV region indicate the average
              EGRET spectrum (Hartman et al. 1999). Solid line reports
              the results of the modelling with the
              synchrotron-inverse Compton model (see text for details
              and model parameters). We also report the single
              emission components: synchrotron (dashed), SSC
              (dotted-dashed) and EC (long dashed).The dotted line
              shows the emission of the accretion disk.}
         \label{sed}
   \end{figure*}

The Spectral Energy Distribution (SED) of 3C454.3 around the epoch of
the July and November 2007 MAGIC observations, assembled with the
available data, is shown in Figure \ref{sed}. For July (left panel) we
show the nearly simultaneous data in the optical (KVA, July 24),
optical-UV ({\it Swift}/UVOT, July 26), X-ray ({\it Swift}/XRT, July 26) and
$\gamma$-ray (AGILE/GRID) band (average of July 24-30). For November
(right panel) the data in the optical, X-ray ({\it Swift}/XRT and INTEGRAL)
and $\gamma$-ray (AGILE/GRID) data, averaged over the entire period of
the AGILE observations (Nov. 11--Dec. 1) are shown.  For comparison we
also show (open circles) historical data. AGILE/GRID spectra, both for
the July and November observations, have been recently published in
Vercellone et al. (2008b).


%

In the same figure, upper limits from MAGIC observations (18-21 July
and 27, 28, 30 November) are shown as triangles (observed: empty;
EBL-deabsorbed: filled) (see Table \ref{tab:ulJulyAugust}). For the
EBL deabsorption we used the LowSFR model of Kneiske et al. (2004)
which predicts a low level of the EBL close to what is presently
inferred from observations, both directly (e.g. Franceschini et
al. 2008) and indirectly (Aharonian et al. 2006b, Mazin \& Raue 2007,
Albert et al. 2008a).

Fig. \ref{sed} shows that the (absorption-corrected) MAGIC upper limit
at $\sim$100 GeV is inconsistent with the extrapolation of the hard
$\gamma$-ray (100 MeV-10 GeV) spectrum. Therefore, the data indicate
there is a break (or a cutoff) of the emission between the GeV and the
100 GeV band. As discussed below this is consistent with the
expectations from the simplest leptonic model.

Emission from blazars is dominated by the non-thermal continuum
emitted by a relativistic jet closely aligned towards the
observer. The SED of FSRQs is widely interpreted in terms of
synchrotron and inverse Compton emission from high-energy
electrons. The latter component is probably dominated by the
scattering of the external photons (originating in the disk and/or in
the broad line region [BLR], Sikora et al. 1994), although the
synchrotron self-Compton emission (Maraschi et al. 1992) and the
inverse Compton scattering of the direct radiation from the accretion
disk (e.g. Dermer \& Schlickeiser 1993) can significantly contribute
in the X-ray band. The SED of 3C454.3, including optical, X-rays and
GeV measurements around the end of July, has been already discussed
and modelled by Ghisellini et al. (2007). However, the model discussed
in that work assumes that the $\gamma$-ray spectrum was similar to the
average EGRET spectrum, with a soft slope. The spectrum of AGILE
(Vercellone et al. 2008b), instead, is rather hard (photon index
$\Gamma \simeq 1.7$ in the 100 MeV-1 GeV band), both for July and November,
suggesting a peak of the high-energy component at frequencies above
$\sim 10^{23}$ Hz. The SEDs of November has been already discussed and
modeled in Vercellone et al. (2008b).

To reproduce the multifrequency data we use the emission model fully
described in Maraschi \& Tavecchio (2003). Given the focus on the VHE
emission, we also consider the absorption of $\gamma$-ray photons
through pair production within the BLR. Moreover, the external
radiation field (assumed to be isotropic in the frame of the black
hole), usually approximated by a black body emission peaking in the UV
region, has been calculated using the photoionization code {\em
CLOUDY} (Ferland et al. 1998). Details on the emission model can be
found in Maraschi \& Tavecchio (2003), while the description of the
calculation of the external radiation field is reported in Tavecchio
\& Ghisellini (2008). 

We assume that the emission is produced within a spherical region of
radius $R$, in motion with bulk Lorentz factor $\Gamma$. We assume
that the corresponding relativistic Doppler factor is $\delta=1/\Gamma
$. The tangled magnetic field has an intensity $B$. The emitting
particles, with density $K$, follow a (steady state) broken-power law
energy distribution extending from $\gamma _{\rm min}$ to $\gamma
_{\rm max}$, with indices $n_1$ and $n_2$ below and above the break at
$\gamma _b$. This {\it purely phenomenological} distribution has been
assumed to reproduce the observed shape of the blazar SEDs, without
any specific assumption on the acceleration/cooling mechanism acting
on the particles. With this choice we are allowed to assume extreme
low-energy slopes ($n_1<2$) such as those required for 3C454.3, which
cannot be obtained under standard conditions. It is conceivable that,
at least in these cases, the electron distribution derives from two
(continuously operating) different acceleration mechanisms (see
e.g. Sikora et al. 2002). We also neglect the effects related to the
cooling of particles in the Klein-Nishina regime, discussed by
Moderski et al. (2005). We note, however, that these effects should
produce a bump in the optical-UV synchrotron emission which is not
apparent in the available data, though the poor coverage does not
allow a firm conclusion. We model the external radiation field
assuming that the disk emission (dotted line in figure), with a total
luminosity of $L_{\rm disk}=5\times 10^{46}$ erg/s, is reprocessed by
clouds of the BLR, a sphere with radius $3\times 10^{17}$ cm (we
assume that clouds are characterized by standard values of the density
$n_{\rm BLR}=10^{11}$~cm$^{-3}$ and hydrogen column density, $N_{\rm
H}=10^{23}$~cm$^{-2}$). For simplicity, we assume that the distance of
the emission region of the jet from the central black hole is smaller
than the radius of the BLR, but large enough to neglect the direct
disk emission, coming from behind the jet (e.g. Dermer \& Schlickeiser 1993;
Vercellone et al. 2008b consider in the model also this
component). To reproduce the shape of the high-energy component, we
assume that in the X-ray band the emission is dominated by SSC
emission, while EC radiation accounts for the GeV peak.

\begin{table}
\begin{center}
\begin{tabular}{ccccccccc}
\hline
$\Gamma$ & $B$ & $K$ & $n_1$& $n_2$& $\gamma _{\rm min}$& $\gamma _b$ & $\gamma _{\rm max}$& $R$\\
\hline
 18.4& 3.1& $5\times 10^5$& 1.9 & 3.6& 85& 500& $6.5\times 10^3$& $6.5$\\
\hline
 17.8& 5& $5\times 10^5$& 1.9 & 3.9& 80& 500& $3.9\times 10^3$& $5$\\
\hline
\end{tabular}
\end{center}
\caption{Parameters used in the emission model to reproduce the SEDs
of 3C454.3 in July (first row) and November (second row)
2007. $\Gamma$: bulk Lorentz factor of the emission region; $B$:
magnetic field, in units of [G]; $K$: density of the relativistic
electrons, in units of [cm$^{-3}$]; $n_1$ and $n_2$: low and high
energy slope of the electron energy distribution; $\gamma _{\rm min}$,
$\gamma _b$ and $\gamma _{\rm max}$: minimum energy, break energy and
maximum energy of the electron energy distribution; $R$: radius of the
emission region, in units of [$10^{15}$ cm]. See text for more
details.}
\label{model}
\end{table}%

In our model, the rapid decrease of the flux above few tens of GeV
would be related to two effects: i) the decrease of the scattering
cross section and ii) the absorption of the produced $\gamma$-rays
through pair production. The energy above which the KN effects become
important can be roughly expressed as: $E_{\rm KN} \simeq 22.5 \nu
_{o,15}^{-1}$ GeV, where $\nu _{o,15}$ is the frequency of the
external photons (in units of $10^{15}$ Hz). The emission including
only the KN effects, neglecting the absorption, is shown by the
long dashed-dotted line. The frequency above which the absorption of
$\gamma$-rays become effective can be roughly expressed as: $E_{\rm
\gamma \gamma} \simeq 60 \nu _{t,15}^{-1}$ GeV, where $\nu _{t,15}$ is
the frequency of the target photons (in units of $10^{15}$
Hz). Therefore, as shown by the solid line in Fig.\ref{sed}
(calculated including both effects, the internal absorption treated as
in Tavecchio \& Mazin 2008), the expected emission above 20-30 GeV is
rather small, consistent with the observed upper limits. Note that,
although the limit set by KN effects is a characteristic feature of
leptonic models, absorption of $\gamma$-rays by soft photons can also
be relevant for hadronic models (e.g. Reimer 2007). More detailed
discussions on the effects of absorption, including the possible
hardening of the observed spectrum above $\sim 100$ GeV, can be found
in, e.g., Aharonian et al. (2008b), Sitarek \& Bednarek (2008),
Tavecchio \& Mazin (2008).

Summarizing, we have shown that the upper limits in the VHE band for
the 3C454.3 obtained with the MAGIC telescope are consistent with the
expectations of the leptonic models for FSRQs, predicting a sharp
decrease of the flux above few tens of GeV, due to the internal
absorption of $\gamma $-rays and the decreased efficiency of the
inverse Compton emission at high energy. Therefore even upper limits,
particularly if accompanied by simultaneous observations in the
MeV-GeV band, can be useful to test current emission models for
FSRQs. Stronger constraints will definitely be obtained with future
multifrequency campaigns already planned with the {\it Fermi}
Gamma-Ray Telescope.

\begin{acknowledgements}
We would like to thank the Instituto de Astrofisica de Canarias for
the excellent working conditions at the Observatorio del Roque de los
Muchachos in La Palma. The support of the German BMBF and MPG, the
Italian INFN and Spanish MCINN is gratefully acknowledged. This work
was also supported by ETH Research Grant TH 34/043, by the Polish
MNiSzW Grant N N203 390834, and by the YIP of the Helmholtz
Gemeinschaft. This research has made use of the NASA/IPAC
Extragalactic Database (NED) which is operated by the Jet Propulsion
Laboratory, California Institute of Technology, under contract with
the National Aeronautics and Space Administration.
\end{acknowledgements}


\begin{thebibliography}{}


\bibitem[]{} Acciari, V.A., et al.\ 2008, ApJ, 685, L73

\bibitem[Aharonian et al.(2008)]{2008RPPh...71i6901A} Aharonian, F.,
Buckley, J., Kifune, T., \& Sinnis, G.\ 2008a, Reports on Progress in
Physics, 71, 096901

\bibitem[Aharonian et al.(2008)]{2008MNRAS.387.1206A} Aharonian, F.~A., 
Khangulyan, D., \& Costamante, L.\ 2008b, MNRAS, 387, 1206 

\bibitem[\protect\citeauthoryear{Aharonian et al.}{2006}]{2006Sci...314.1424A} 
Aharonian F., 
et al., 2006a, Science, 314, 1424

\bibitem[Aharonian et al.(2006)]{2006Natur.440.1018A} Aharonian, F., et 
al.\ 2006b, Nature, 440, 1018 

\bibitem[\protect\citeauthoryear{Aharonian et 
al.}{2003}]{2003A&A...403L...1A} Aharonian F., et al., 2003, A\&A, 403, L1 

\bibitem[]{} Albert, J. et al., 2007, ApJ, 666, L17

\bibitem[MAGIC Collaboration et al.(2008)]{2008Sci...320.1752M}
Albert, J. et al.\ 2008a, Science, 320,1752

\bibitem[Albert et al.(2008)]{2008ApJ...674.1037A} Albert, J., et al.\ 
2008b, ApJ, 674, 1037

\bibitem{} Albert, J. et al., 2008c, Nucl. Instr. Meth. A, 558, 424

\bibitem{} Albert, J. et al., 2008d, Astropart. Phys., in press
(arXiv:0810.3586)

\bibitem[]{} Aliu, E.. et al., 2008, ApJ, submitted

\bibitem[]{} Baixeras, C., et al.\ 2004, Nuclear Instruments and
Methods in Physics Research A, 518, 188

\bibitem[]{rf} Breiman, L. 2001, Machine Learning, 45, 5

\bibitem[Bretz \& et al.(2005)]{2005ICRC....4..315B} Bretz, T., \& et
al.\ 2005, in: Proc. 29th Int. Cosm. Ray Conf. (Pune, India), 4, 315

\bibitem[Cortina \& et al.(2005)]{2005ICRC....5..359C} Cortina, J., \&
et al.\ 2005, International Cosmic Ray Conference, 5, 359

\bibitem[]{} De Angelis, A., Mansutti, O., Massimo, P., 2008, La
Rivista del Nuovo Cimento, 31, n.4, 187 (arXiv:0712.0315)

\bibitem[Dermer \& Schlickeiser(1993)]{1993ApJ...416..458D} Dermer,
C.~D., \& Schlickeiser, R.\ 1993, ApJ, 416, 458

\bibitem[]{} Fegan, D. J., 1997, J. Phys. G, 23, 1013

\bibitem[\protect\citeauthoryear{Ferland et
al.}{1998}]{1998PASP..110..761F} Ferland G.~J., Korista K.~T., Verner
D.~A., Ferguson J.~W., Kingdon J.~B., Verner E.~M., 1998, PASP, 110,
761

\bibitem[]{wobble} Fomin, V.~P., Stepanian, A., A., Lamb, R.~C., Lewis,
D.~A., Punch, M., \&   Weekes, T.~C. 1994, Astroparticle Physics, 2, 137

\bibitem[Franceschini et al.(2008)]{2008A&A...487..837F} Franceschini,
A., Rodighiero, G., \& Vaccari, M.\ 2008, A\&A, 487, 837

\bibitem[Gear et al.(1994)]{1994MNRAS.267..167G} Gear, W.~K., et al.\ 1994, 
MNRAS, 267, 167 

\bibitem[]{} Giommi P. et al., 2006, A\&A, 456, 911

\bibitem[Ghisellini et al.(2007)]{2007MNRAS.382L..82G} Ghisellini, G., 
Foschini, L., Tavecchio, F., \& Pian, E.\ 2007, MNRAS, 382, L82 

\bibitem[]{mux} Goebel F., et al. in: Proc. 30th Int. Cosm. Ray
Conf. (Merida, Mexico), preprint: arXiv:0709.2605

\bibitem[Hartman et al.(1999)]{1999ApJS..123...79H} Hartman, R.~C., et al.\ 
1999, ApJS, 123, 79 

\bibitem[]{}Hillas, A. M., 1985, Proc. of the 19th ICRC, La Jolla, 3, 445 

\bibitem[Kneiske et al.(2004)]{2004A&A...413..807K} Kneiske, T.~M.,
Bretz, T., Mannheim, K., \& Hartmann, D.~H.\ 2004, A\&A, 413, 807

\bibitem[Kuehr, Pauliny-Toth, Witzel, \&
Schmidt(1981)]{1981AJ.....86..854K} Kuehr, H., Pauliny-Toth, I.\ I.\ K.,
Witzel, A., \& Schmidt, J.\ 1981, AJ, 86, 854

\bibitem[Impey \& Neugebauer(1988)]{1988AJ.....95..307I} Impey, C.\ D.\ \& 
Neugebauer, G.\ 1988, AJ, 95, 307 

\bibitem[\protect\citeauthoryear{Liu \& Bai}{2006}]{2006ApJ...653.1089L} 
Liu H.~T., Bai J.~M., 2006, ApJ, 653, 1089 

\bibitem[Maraschi \& Tavecchio(2003)]{2003ApJ...593..667M} Maraschi,
L., \& Tavecchio, F.\ 2003, ApJ, 593, 667

\bibitem[Maraschi et al.(1992)]{1992ApJ...397L...5M} Maraschi, L., 
Ghisellini, G., \& Celotti, A.\ 1992, ApJ, 397, L5 

\bibitem[Mazin \& Raue(2007)]{2007A&A...471..439M} Mazin, D., \& Raue,
M.\ 2007, A\&A, 471, 439

\bibitem[Moderski et al.(2005)]{2005MNRAS.363..954M} Moderski, R., Sikora, 
M., Coppi, P.~S., \& Aharonian, F.\ 2005, MNRAS, 363, 954

\bibitem[]{} Pian E. et al., 2006, A\&A, 449, L21

\bibitem[\protect\citeauthoryear{Reimer}{2007}]{2007ApJ...665.1023R} Reimer 
A., 2007, ApJ, 665, 1023

\bibitem[]{rolke} Rolke, W.~A., L\'{o}pez, A.~M., \& Conrad, J. 2005,
Nucl. Instr. Meth., A551, 493

\bibitem[\protect\citeauthoryear{Sikora, Begelman, \&
Rees}{1994}]{1994ApJ...421..153S} Sikora M., Begelman M.~C., Rees
M.~J., 1994, ApJ, 421, 153

\bibitem[Sikora et al.(2002)]{2002ApJ...577...78S} Sikora, M., 
B{\l}a{\.z}ejowski, M., Moderski, R., \& Madejski, G.~M.\ 2002, ApJ, 577, 
78 

\bibitem[Sitarek \& Bednarek(2008)]{2008arXiv0807.4228S} Sitarek, J.,
\& Bednarek, W.\ 2008, MNRAS, in press (arXiv:0807.4228)

\bibitem[Smith et al.(1988)]{1988ApJ...326L..39S} Smith, P.\ S., Elston, 
R., Berriman, G., Allen, R.\ G., \& Balonek, T.\ J.\ 1988, ApJ, 326, L39 

\bibitem[Stevens et al.(1994)]{1994ApJ...437...91S} Stevens, J.\ A., 
Litchfield, S.\ J., Robson, E.\ I., Hughes, D.\ H., Gear, W.\ K., 
Terasranta, H., Valtaoja, E., \& Tornikoski, M.\ 1994, ApJ, 437, 91 

\bibitem[]{}Swordy, S. et al., 2008, The Astronomer's Telegram, 1753

\bibitem[]{}Tavani, M. et al. 2008, A\&A, submitted, (arXiv:0807.4254)

\bibitem[]{}Tavecchio, F., Mazin, M., 2008, MNRAS, in press
(arXiv:0809.2467)

\bibitem[Tavecchio et al.(2007)]{2007ApJ...662..900T} Tavecchio, F., 
Maraschi, L., Wolter, A., Cheung, C.~C., Sambruna, R.~M., 
\& Urry, C.~M.\ 2007, ApJ, 662, 900 

\bibitem[Tavecchio et al.(2002)]{2002ApJ...575..137T} Tavecchio, F.,
Ghisellini, G., 2008, MNRAS, 386, 945

\bibitem[Tavecchio et al.(2002)]{2002ApJ...575.37T} Tavecchio, F.,
Mazin, D., 2008, MNRAS, submitted (arXiv:0809.2467)

\bibitem[]{diego} Tescaro D., et al., in: Proc. 30th Int. Cosm. Ray
Conf. (Merida, Mexico), preprint: arXiv:0709.1410

\bibitem[]{}Teshima, M. et al. 2008, The Astronomer's Telegram, 1500

\bibitem[Vercellone et al.(2008)]{2008ApJ...676L..13V} Vercellone, S., et 
al.\ 2008a, ApJ, 676, L13 

\bibitem[Vercellone et al.(2008b)]{2008ApJ...676L..3V} Vercellone, S., et 
al.\ 2008b, ApJ, in press, (arXiv:0809.1737)

\bibitem[]{} Villata M. et al., 2006, A\&A, 453, 817


\end{thebibliography}
\end{document}